\documentclass{ifacconf}

\usepackage{multirow}

\usepackage{stfloats}

\usepackage{amsmath}
\usepackage{amssymb}
\usepackage{enumerate}
\usepackage{definice}
\usepackage{color}
\usepackage{booktabs,hhline} 
\usepackage{cool}
\usepackage{natbib}
\usepackage{mathtools}

\usepackage[scr=boondoxo,scrscaled=1.05]{mathalfa}

\def\bfA{\mathbf A}

\def\bfD{\mathbf D}

\def\bfF{\mathbf F}
\def\bfG{\mathbf G}

\def\bfI{\mathbf I}

\def\bfP{\mathbf P}
\def\bfQ{\mathbf Q}
\def\bfR{\mathbf R}
\def\bfS{\mathbf S}
\def\bfT{\mathbf T}

\def\bfg{\mathbf g}

\def\bfr{\mathbf r}

\def\bfv{\mathbf v}
\def\bfw{\mathbf w}
\def\bfx{\mathbf x}
\def\bfy{\mathbf y}
\def\bfz{\mathbf z}

\def\bfDelta{\mathbf \Delta}

\def\bfLambda{\mathbf \Lambda}

\DeclareMathAlphabet\mathbfcal{OMS}{cmsy}{b}{n}

\def\real{\mathbb{R}}

\def\T{^T}

\DeclarePairedDelimiter\ceil{\lceil}{\rceil}

\usepackage{graphicx}      
\usepackage{natbib}        

\newcommand{\tightoverset}[2]{%
	\mathop{#2}\limits^{\vbox to -.5ex{\kern-0.75ex\hbox{$#1$}\vss}}}

\begin{document}
	\begin{frontmatter}
		
		\title{\hspace{-1mm}Efficient~Point~Mass~Predictor~for~Continuous and Discrete Models with Linear Dynamics\hspace{-0mm}}
		
		\thanks[footnoteinfo]{This work was supported by the Czech Science Foundation (GACR) under grant GA 22-11101S.}
		
\author[First]{J. Matou\v{s}ek} 
\author[First]{J. Dun\'{i}k} 
\author[Second]{M. Brandner}
\author[Third]{Chan Gook Park} 
\author[Fourth]{Yeongkwon Choe} 

\address[First]{Dept. of Cybernetics, University of West Bohemia, Pilsen, Czech Republic (e-mails: \{matoujak,dunikj\}@kky.zcu.cz).}
\address[Second]{Dept. of Mathematics, University of West Bohemia, Pilsen, Czech Republic (e-mail: brandner@kma.zcu.cz)}
\address[Third]{Dept. of Aerospace Eng. and Automation and Systems Research Institute, Seoul National University, Seoul, Korea (e-mail: chanpark@snu.ac.kr)}
\address[Fourth]{Mobility Platform Research Center, Korea Electronics Technology Institute, Gyeonggi-do, Korea (e-mail: veritasbbo@gmail.com)}
		
		\begin{abstract} 
		This paper deals with state estimation of stochastic models with linear state dynamics, continuous or discrete in time. The emphasis is laid on a numerical solution to the state prediction by the time-update step of the grid-point-based point-mass filter (PMF), which is the most computationally demanding part of the PMF algorithm. A novel way of manipulating the grid, leading to the time-update in form of a convolution, is proposed. This reduces the PMF time complexity from quadratic to log-linear with respect to the number of grid points. Furthermore, the number of unique transition probability values is greatly reduced causing a significant reduction of the data storage needed. The proposed PMF prediction step is verified in a numerical study.
		\end{abstract}
		
		\begin{keyword}
			State estimation, prediction, transition probability matrix, Chapman-Kolmogorov equation, Fokker-Planck equation, point-mass filter, convolution.
		\end{keyword}
		
	\end{frontmatter}

	\section{Introduction}
	State estimation deals with computing unknown values from in-directly related and noisy data and a mathematical model of the considered system. Estimation algorithms are crucial components of any modern signal processing and fault detection system in areas ranging from navigation and automatic control through weather forecasting to medical applications.
	
	A general solution to the state estimation problem is given by the Bayesian recursive relations (BRRs). These relations describe the evolution of the probability density functions (PDFs) of the state conditioned on the measurements by the Bayes' rule for the measurement update and the Chapman-Kolmogorov equation (CKE) or the Fokker-Planck equation (FPE) for the time update. The CKE is used when the time evolution of the state is modeled by the discrete dynamics (DD) model, whereas the FPE is when the continuous dynamics (CD)  model is at our disposal. Although the conditional PDF provides a full description of the immeasurable state of a nonlinear stochastic dynamic system, the relations are tractable exactly for a limited set of models only, where the assumption on linearity or Gaussianity is usually considered. As an example of an exact estimator, the Kalman (Bucy) filter (KF)  can be mentioned. For other models, the recursive relations are solved approximately \citep{Sa:13}.
	
	In this paper, a numerical solution to the BRRs using the point-mass filter (PMF) is considered with a particular emphasis on the prediction (or time-update) step  \citep{Be:99,SiKraSo:06,JePaPa:18}. In the prediction step of the PMF, the integral CKE or the partial differential FPE are solved using \textit{deterministic} integration rules or numerical schemes, respectively. A significant region of the state-space is covered by the grid of points in which the conditional PDF is computed. This numerical solution is the most computationally expensive part of the PMF with complexity growing \textit{exponentially} with state-space dimension. Therefore, a number of methods have been proposed to reduce the PMF computational complexity, but often at the costs of additional approximations or strong requirements \citep{SmGa:13,DuSoVeStHa:19,Be:99,DuStMaBl:22}.
	
	This paper presents an approach for the design of the computationally \textit{efficient} point-mass prediction (PMP) with \textit{minimal memory} requirements. The approach is based on such a definition of the predictive grid, which allows the usage of a convolution theorem, and leads to the computation of a part of the state transition probability matrix only. As a consequence, the proposed PMP, for state-space models with discrete and continuous dynamics, has a \textit{log-linear} complexity and linearly increasing data storage need with respect to the number of grid points.

	\section{Model and Point-Mass State Prediction}
	The PMP is based on a grid of points covering a significant part of the state-space sufficiently well. Therefore, in this section, the conditional PDF approximation on a grid is introduced first followed by a review of standard solutions for state prediction for the DD and CD models with the stress on their computational complexity.
	
	\subsection{Point Mass Density Approximation}
	
\begin{figure}
		\includegraphics[width=0.5\textwidth]{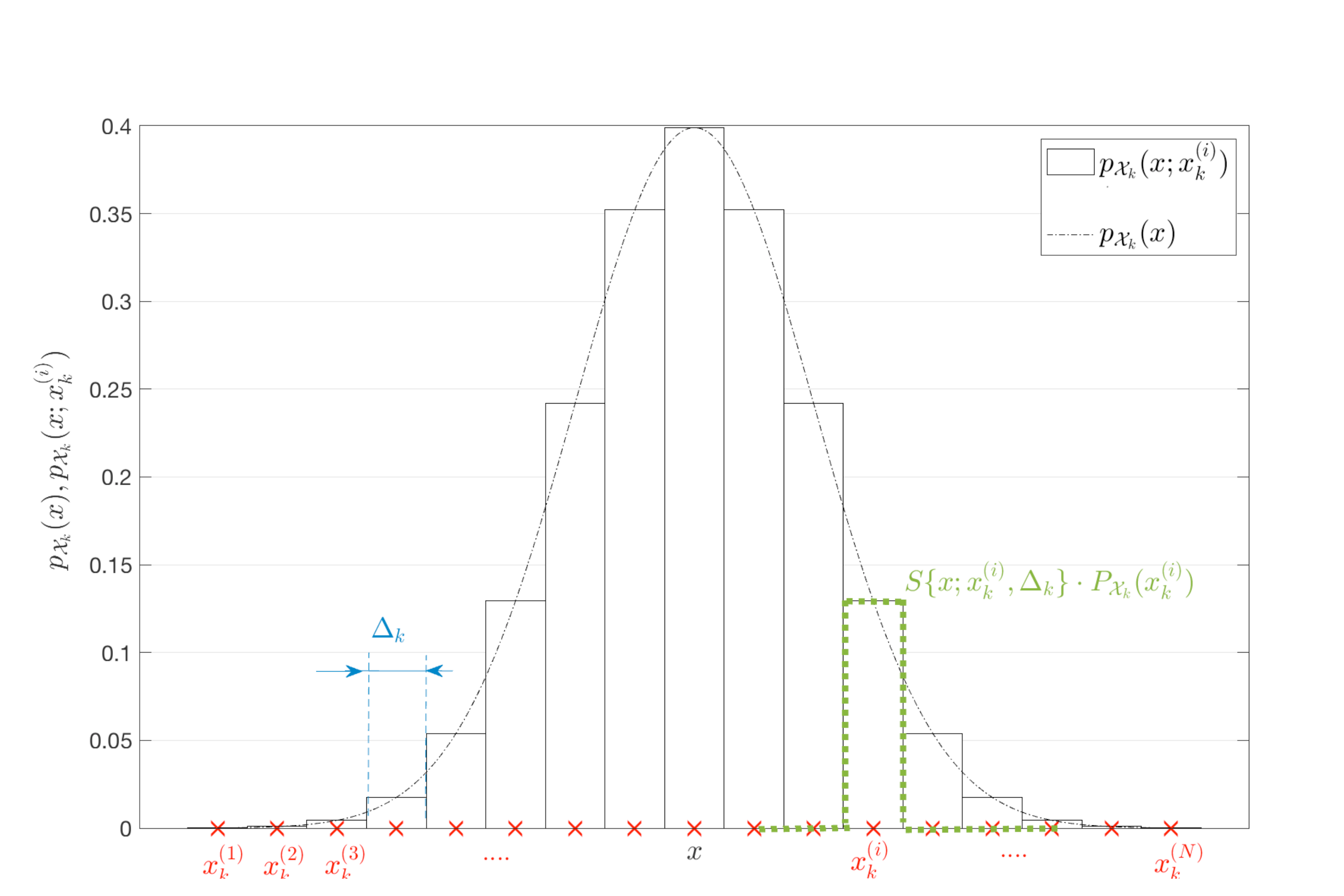}
		\caption{Point mass density illustration}
		\label{fig:PMD}
	\end{figure}

	The PMP for both DD and CD models is based on an approximation of a (conditional) PDF $p_{\mathbfcal{X}_k}(\bfx)$ by a \textit{piece-wise constant} point-mass density (PMD) $p_{\mathbfcal{X}_k}(\bfx;\bfx_k^{(:)})$ computed at the set of $N$ ``discrete'' grid points $\bfx_k^{(:)}=\{\bfx_k^{(i)}\}_{i=1}^N , \bfx^{(i)}_k\in\real^{n_x}$, as follows \citep{SiKraSo:06}
	\begin{align}
	p_{\mathbfcal{X}_k}(\bfx;\bfx_k^{(:)})\triangleq\sum_{i=1}^NP_{k}^{(i)}S_{\mathbfcal{X}_k}\{\bfx;\bfx^{(i)}_k,\bfDelta_k\},\label{eq:pdf_pm}
	\end{align}
	where
	\begin{itemize}
		\item $\bfx$ is the state space variable realization,
		\item $\mathbfcal{X}_{k}: \Omega \rightarrow \real^{n_{\bfx}}$ is the random state variable at discrete time instant $k$,
		\item  $N = N_1 \cdot N_2\ ... \cdot N_{n_x}$ and $N_i$ is a number of discretisation steps per marginal PDF of $i$-th state vector element, $N_{pa}$ is be used, if $N_1 = N_2 = ... = N_{n_x} = N_{pa}$, 
		\item $P_{k}^{(i)}=c_k\tilde{P}_{k}^{(i)}$, where $\tilde{P}_{k}^{(i)}=p_{\mathbfcal{X}_k}(\bfx;\bfx_k^{(i)})$ is the value of the PDF $p_{\mathbfcal{X}_k}(\bfx)$ evaluated at the $i$-th grid point $\bfx^{(i)}_k$ further also called as a \textit{weight}, $c_k=\delta_k\sum_{i=1}^{N}\tilde{P}_{k}^{(i)}$ is a normalisation constant, and $\delta_k$ is the volume of the $i$-th point neighbourhood defined as its a (hyper-)rectangular neighbourhood
		\begin{align}
		\bfDelta_k=[\bfDelta_{k,1}, \bfDelta_{k,2}, \ldots, \bfDelta_{k,{n_x}}]\T,
		\end{align}	
		where the PMD $p_{\mathbfcal{X}_k}(\bfx;\bfx_k^{(:)})$ is constant and has value $P_{\mathbfcal{X}_k}^{(i)}$, and
		\item $S_{\mathbfcal{X}_k}\{\bfx;\bfx^{(i)}_k,\bfDelta_k\}$ is a \textit{selection} function defined as
		\begin{align}
		S_{\mathbfcal{X}_k}\{\!\bfx;\bfx^{(i)}_k\!,\!\bfDelta_k\!\}\!=\!\begin{cases}
		\!1,\mathrm{if}\ |\bfx\!-\!\bfx^{(i)}_k|\!\leq\!\tfrac{\bfDelta_k}{2} \text{ per element},\\
		\!0, \mathrm{otherwise}.
		\end{cases}\label{eq:sf3}
		\end{align}
		
	\end{itemize}
	For convenience, the vector of all weights associated with grid points $\bfx_k^{(:)}$ is further noted as
	\begin{align}
	P_{k}^{(:)} \overset{\Delta}{=}	
	\begin{bmatrix}
	P_{k}^{(1)}\\
	\vdots\\
	P_{k}^{(N)}\\
	\end{bmatrix}.
	\end{align}
	An illustration of a PMD approximating a PDF is drawn in Fig.~\ref{fig:PMD}. PMD properties and moment computations can be found e.g., in \citep{Be:99}.

	\subsection{Discrete Dynamics Model and State Prediction}\label{DDSE}
	The DD model of the state is given by the stochastic difference equation
	\begin{align}
	\mathbfcal{X}_{{k+1}}&=\bfF \mathbfcal{X}_{k} + \mathbfcal{W}_{k},\label{eq:asx} k=0,1,2,\ldots,T,
	\end{align}
	where $\mathbfcal{X}_{k}: \Omega \rightarrow \real^{n_{\bfx}}$ is the state random variable at discrete time instant $k$ of which realisation $\bfx$ is \textit{unknown} and \textit{estimated}, $\bfF \in \real^{n_{\bfx} \times n_{\bfx}}$ is a \textit{known} matrix, $\Omega$ is the event space, and $\mathbfcal{W}_{k}$ is a white noise with \textit{known} time-invariant PDF $p_{\mathbfcal{W}_{k}}(\bfw)$. The initial condition PDF $p_{\mathbfcal{X}_{0}}(\bfx)$ is assumed to be known. The initial state is independent of the state noise. 
	
	
	The general solution to the state prediction is given by the CKE \citep{AnMo:79}
	\begin{align}
	p_{\mathbfcal{X}_{k+1}}(\bfx)&=\int p_{\mathbfcal{X}_{k+1}|\mathbfcal{X}_{k}}(\bfx|\bfy)p_{\mathbfcal{X}_k}(\bfy)d^{n_{\bfx}}\bfy,\label{eq:pred}
	\end{align}
	where $\bfx,\bfy \in \real^{n_{\bfx}}$ are the state variable realizations, $p_{\mathbfcal{X}_{k+1}|\mathbfcal{X}_{k}}(\bfx|\bfy) = p_{\mathbfcal{W}}(\bfx - \bfF\bfy)$ is the known state transition PDF obtained from \eqref{eq:asx}, and $p_{\mathbfcal{X}_k}$ is the last available PDF of the state $\mathbfcal{X}_k$ at time $k$, further called \textit{initial}. This variable can come from the previous prediction or filtering step or from the initial condition $p_{\mathbfcal{X}_{0}}$. The sought one-step \textit{predictive} PDF at time $k+1$ is denoted as $p_{\mathbfcal{X}_{k+1}}$.
	
	
	The numerical solution to the CKE starts from the assumption of the initial PDF in the form of PMD and the creation of the ``new'' grid $\bfx_{k+1}^{(:)}$ at subsequent time $k+1$. The new grid location is determined on the basis of easily computable (by a local filter) predictive moments. 
	Then, the basic\footnote{Basic means that the standard mid-point integration rule is employed.} point-mass solution to the CKE \eqref{eq:pred} at the new grid $\bfx_{k+1}^{(:)}$ reads \citep{SiKraSo:06}
	\begin{align}
	p_{\mathbfcal{X}_{k+1}}(\bfx;\bfx_{k+1}^{(:)})=\sum_{j=1}^N P_{{k+1}}^{(j)}S_{\mathbfcal{X}_{k+1}}\{\bfx;\bfx^{(j)}_{k+1},\bfDelta_{k+1}\},\label{eq:pdf_pm_pred}
	\end{align} 
	where the value of the predictive PDF at $j$-th grid point is computed by
	\begin{align}
	P_{{k+1}}^{(j)}\!=\!\sum_{i=1}^N p_{\mathbfcal{X}_{k+1}|\mathbfcal{X}_{k}}(\bfx^{(j)}_{k+1}|\bfx^{(i)}_{k})P_{k}^{(i)}\delta_k.\label{eq:pdf_predII}
	\end{align} 
	Alternatively, the equation \eqref{eq:pdf_predII} can be conveniently written in a matrix form as
	\begin{align}
	P_{{k+1}}^{(:)} =\bfF_{\text{trans}}\ P_{{k}}^{(:)}, \label{eq:CKEnumSol}
	\end{align}
	where $\bfF_{\text{trans}}\in\real^{N\times N}$ is an matrix with the element in $j$-th row and $i$-th column given by
	\begin{align}
	\bfF_{\text{trans},j,i} &= p_{\mathbfcal{X}_{k+1}|\mathbfcal{X}_{k}}(\bfx^{(j)}_{k+1}|\bfx^{(i)}_k)\delta_k\nonumber\\
	&=p_{\mathbfcal{W}_k}\left(\bfx^{(j)}_{k+1} -  \bfF\bfx^{(i)}_k\right)\delta_k. \label{eq:TPMmat}
	\end{align}
	
	\subsection{Continuous Dynamics Model and State Prediction}\label{CDSE}
	The CD model is given by the state stochastic differential equation
	\begin{align}
	d\mathbfcal{X}(t)&= \bfA \mathbfcal{X}(t) dt+\bfQ d\bf\mathbfcal{W}(t)\label{eq:dynam},
	\end{align}
	where $\mathbfcal{X}(t): \Omega \rightarrow \real^{n_{\bfx}}$ is the state random variable at (continuous) time $t$, with $\Omega$ being the event space, $\bfA \in\real^{n_{\bfx} \times n_{\bfx}}$ is the \textit{known}  matrix, $\mathbfcal{W}(t)$ is the state noise modelled by the Brownian motion with \textit{normally} distributed increments with the covariance matrix $E[d\bfw(t)(d\bfw(t))^T]=\bfI_{n_{\bfx}}dt$, and $\bfQ\in\real^{n_{\bfx}\times n_{\bfx}}$ is the \textit{known} matrix of diffusion coefficients. The state noise is independent of the initial state with the \textit{known} PDF $p_{\mathbfcal{X}(0)}(\bfx)$. For consistency with the DD, a notation $\bfx(t_k) \triangleq \bfx_k$ is used throughout this paper. For simplicity, we assume that the sampling period is $t_{k+1} - t_{k} = 1$.
	
	The time evolution of the sought (conditional) PDF is governed by the FPE
	\begin{align}
	{\pderiv{p_{\mathbfcal{X}(t)}(\bfx)}{t}} &= -{\nabla\cdot \left(\bfA \bfx\ p_{\mathbfcal{X}(t)}(\bfx) \right)}\nonumber\\
	& + \frac{1}{2}\nabla\cdot\left(\bfQ\left(\nabla^T p_{\mathbfcal{X}(t)}(\bfx)\right) \right), \label{eq:fokker}
	\end{align}
	where $t\in(k,k+1)$, $\nabla$ denotes the gradient operator as a row vector, and $\nabla \cdot$ is the divergence. In  \eqref{eq:fokker}, the first right-hand side term is named \textit{hyperbolic} and it describes the \textit{advection} of the PDF tied to the state dynamics. The second term is named \textit{parabolic} and it describes the \textit{diffusion} caused by the state noise. Contrary to the CKE \eqref{eq:pred}, the FPE \eqref{eq:fokker} holds for the Gaussian state noise only. For a non-Gaussian noise, the PDF evolution equation would have an infinite number of terms \citep{DuSpa:05}. 
	
	Assuming the initial PDF $p_{\mathbfcal{X}(t_k)}$ in the form of the PMD, the FPE is solved by numerical methods \citep{Pi:13},
	typically by the linear ones \citep{Ka:18,ChaSh00}
	with a numerical method time step $\Delta t \ll (t_{k+1}-t_{k})$. The linear methods can be defined using a matrix $\bfF_{\text{diff}} \in \mathbb{R}^{N \times N}$ \citep{leveque_2002},
	which allows computing the predictive weights (i.e., values of the PMD at the next numerical time step $t_k+\Delta t$)  as
	\begin{align}
	P_{{t_k+\Delta t}}^{(:)} = \bfF_{\text{diff}}(t_k)P_{{t_k}}^{(:)},\label{eq:diffTPM}
	\end{align}
	where $P_{{t_k}}^{(:)}=P_{{k}}^{(:)}$ and $\bfF_{\text{diff}}(t_k)$ is a multi-diagonal numerical step dependent sparse matrix, which is discussed later. The sparse matrix depends on time as it contains terms $\bfA \bfx^{(i)}_k$, that are changing as the grid $\bfx_k^{(:)}$ moves to cover the "important" part of the state space \citep{LiPe:97}. 
	

	The resulting prediction, i.e., a numerical solution to the FPE \eqref{eq:fokker}, from the time instant $k$ to $k+1$, \eqref{eq:diffTPM} becomes
	\begin{align}
	P_{{t_{k+1}}}^{(:)} =\underbrace{\bfF_{\text{diff}}(t_k+l\Delta_t) \cdots \bfF_{\text{diff}}(t_k+\Delta_t)\bfF_{\text{diff}}(t_k)}_{\bfT}\ P_{{t_k}}^{(:)}, \label{eq:FKEnumSol}
	\end{align}
	where $l = \frac{t_{k+1} - t_k}{\Delta t} $ is a power.
	
	
	\subsection{Transition Probability Matrix and its Computational Complexity}
	Both solutions, i.e., the solution in the discrete and continuous time domain, computing the weights of the predictive PMD can be written as a linear matrix equation
	\begin{align}
	P_{{k+1}}^{(:)} =\bfT\ P_{{k}}^{(:)}, \label{eq:TPM}
	\end{align}
	where $\bfT$ is the transition probability matrix (TPM) being equal to $\bfF_{\text{trans}}$ in \eqref{eq:CKEnumSol} or approximately\footnote{The matrix $\bfF_{\text{diff}}^l$ in \eqref{eq:FKEnumSol} is affected by the time-stepping error, this error is further neglected, as it gets smaller with a time-step $\Delta t$, which now has a small performance impact.} equal to $\bfF_{\text{diff}}^l$ in \eqref{eq:FKEnumSol} for DD or CD models, respectively. Independently of the time domain, the TPM calculation is the most computationally demanding operation not only of the PMP but also of the whole PMF. The TPM calculation complexity grows \textit{quadratically} with the number of points $N$, which grows exponentially with the state dimension $n_x$. It means that the PMF prediction complexity is $\mathcal{O}\left( N^2 \right)$, where $N=N_{pa}^{n_x}$. In the discrete case,  it is caused by the need for evaluation of the state noise PDF for all combinations of points $ \bfx^{(:)}_k$ and  $\bfx^{(:)}_{k+1}$. In the continuous case, the complexity lies in the calculation of particular transition matrices and their multiplications.

	\section{Standard PMP Complexity and Goal}\label{sec:mot}
	Construction of the TPM $\bfT$ and calculation of time-update \eqref{eq:TPM} is the most computationally demanding part of the PMF for both DD and CD models. To illustrate its complexity, let the estimation of a five-dimensional state be considered with the number of grid points per dimension $N_{pa} = 11$. That is the grid at one time instant has $N = 11^5 = 161,051$ points. The standard predictive step of the PMF for the DD models \eqref{eq:pdf_pm_pred} then evaluates the transition probability $p_{\mathbfcal{X}_{k+1}|\mathbfcal{X}_{k}}(\bfx|\bfy)$ $N^2 = 25,937,424,601$ times (all combinations of $N$ points at two subsequent time instants), and also calculates $N^2$ operations in the consequent matrix product for the CD model. Therefore at the currently available hardware, the standard PMF is usable for $n_{\bfx} \leq 3$.
	
	\subsection{Computational Complexity Reduction}
	Therefore, a range of techniques for the PMF computational complexity reduction was proposed, however, at the cost of additional approximations, the need for user/designer defined parameters, or for models of a special form. Namely, the following computation reduction techniques can be mentioned:
	\begin{itemize}
		\item \textit{Rao-Blackwellisation} is designed for the conditionally linear structure with Gaussian noises, where the nonlinearly modeled part of the state is estimated by the expensive PMF, whereas the remaining linearly modeled part is estimated by computationally cheap KFs \citep{SmGa:13,DuSoVeStHa:19}.
		\item \textit{Separable prediction} takes advantage of an off-line calculation of the TPM assuming a random walk model (i.e., $\bfF = \bfI$) with a known shift vector, and Gaussian noise $\mathbfcal{W}$. It does not use the convolution theorem. The complexity is dependent on the state noise variance. For the worst case scenario, it has still nearly $\mathcal{O}(N^2)$ \citep{Be:99}. 
		\item \textit{Copula prediction} is based on the propagation of the $n_x$ marginal PDFs and a copula, capturing the correlation, rather than on the propagation of $n_x$ dimensional conditional PDF. An optimal copula cannot generally be found and, thus, its selection is a designer decision leading to an approximation error \citep{DuStMaBl:22}.
		\item \textit{Tensor-based prediction} decomposes the transition PDF into a lower-dimensional set of functional tensors of form determined by the designer. Decomposition inherently leads to an approximation error \citep{TiStDu:22,LiWaYaZh:19}.
	\end{itemize}
	
	Note also that, the PMF grid can be designed in an adaptive or sparse layout for both models. Although, those layouts lead to the reduction of the total number of grid points $N$, the order of the complexity is still $\mathcal{O}\left( N^2 \right)$ \citep{KaSch13}.
	
	
	\subsection{Goal of the Paper}
	The goal of this paper is to propose a computationally and memory-efficient PMP, employing the fast Fourier transform-based convolution. By smart selection of the predictive grid of points, only one row of the TPM has to be computed and stored for given matrices $\bfF$ in \eqref{eq:asx} and  $\bfA$ in \eqref{eq:dynam} of \textit{arbitrary} structure. 
	As a consequence, the computational complexity of the PMF prediction is reduced from \textit{quadratic} into \textit{log-linear} with respect to the number of points $N$.
	

	\section{Efficient Point-Mass Prediction}
	The standard PMP computes the ``new'' grid $\bfx_{k+1}^{(:)}$ for \eqref{eq:pdf_pm_pred} based on the predicted mean and covariance matrix of the state computed simply using a suitable state estimator providing predictive moments \citep{SiKraSo:06}.  In this case, the TPM does not have a diagonal form, and thus cannot be rewritten as a convolution and has to be calculated the usual inefficient way.
	
	\subsection{Main Idea}
	
	The \textit{main} idea of the proposed efficient PMP lies, instead of creating the ``new'' grid on the basis of the first two predictive moments, the grid $\bfx_{k+1}^{(:)}$ is created by transforming the ``old'' $\bfx_{k}^{(:)}$ via the DD or CD dynamics. Then, the time-update step becomes convolution, allowing efficient implementation using the convolution theorem. 
	
	
	\begin{prop}
		Let the ``new'' grid be constructed as 
		\begin{align}
		\bfx_{k+1}^{(i)} &= \bfF\bfx_{k}^{(i)}, \forall i,\label{eq:newGridDD}
		\end{align}
		for the DD model, and 
		\begin{align}
		\mathbf{{x}}^{(i)}_{t_{k+\Delta_t}} &= \exp(\bfA \Delta_t) \mathbf{x}^{(i)}_{t_k}, \forall i,\label{eq:forcMovGrid}
		\end{align}
		for the CD model.
		Let the number of grid points per $i$-th dimension $N_i$ be odd $\forall i$.
		
		For the grids $\bfx_{k+1}^{(:)}$, $\bfx_{k}^{(:)}$, and a DD or CD model, compute the $m$-th TPM row $\bfT_{m,:}$ corresponding to the middle point of the predictive grid $ \bfx_{k+1}^{(m)}$ only, where $m = \ceil*{\frac{N}{2}}$, and reshape it to be aligned with the physical grid space (this is illustrated in subsection \ref{subSec:effForDD}) leading to 
		\begin{align}
		\widetilde\bfT_{m,:} = \psi \left( \bfT_{m,:} \right) \in \mathbb{R}^{N_1 \times\ldots\times N_{n_x}},\label{eq:tmdp_aligned}
		\end{align}
		with an appropriately sorted grid of points and the reshaped initial PMD weights $\widetilde{P}_{{k}}^{(:)}$ and $\psi : \mathbb{R}^{N \times 1} \rightarrow \mathbb{R}^{N_1 \times\ldots\times N_{n_x}}$ being the reshape operator. Then, the predictive weights can be calculated by $n_x$-dimensional convolution with zero padding as
		\begin{align}
		\widetilde{P}_{{k+1}}^{(:)} = \widetilde\bfT_{m,:} * \overset{n_x}{...} *  \widetilde{P}_{{k}}^{(:)}, \label{eq:tpmconv}
		\end{align}
		where the symbol $* \overset{n_x}{...} *$ denotes the convolution in $n_x$-dimensional space.

		Now the convolution theorem can be applied to \eqref{eq:tpmconv}, for efficient calculation in the frequency domain
		\begin{align}
		\widetilde P_{{k+1}}^{(:)} = \mathcal{F}^{-1}\left(  \mathcal{F}(\widetilde{\bfT}_{m,:}) \odot \mathcal{F}(\widetilde{ P}_{{k}}^{(:)})\right), \label{eq:tpmconvFFT}
		\end{align}
		where $\mathcal{F}$ denotes the Fourier transform and $\odot$ the Hadamard product. 
	\end{prop}

	With respect to the dimensionality of the considered matrices, the efficient PMP evaluating \eqref{eq:tpmconvFFT} has log-linear complexity $\mathcal{O}( N \log N )$.

	
	Note that, the proposed prediction procedure's ``new'' grid does not respect the state noise properties, i.e., the noise covariance matrix $\bfQ$ is ignored. To take into account the contribution of the state noise on the ``new'' grid layout, the grid can be adjusted as long as the points are equidistantly spaced i.e., the posterior grid can be made bigger and new posterior PMD weights interpolated/extrapolated in order for the new grid \eqref{eq:newGridDD}, \eqref{eq:forcMovGrid} respecting the state noise magnitude.
	
	\subsection{Efficient PMP for DD Model}\label{subSec:effForDD}
\begin{figure*}
		\includegraphics[width=1\textwidth]{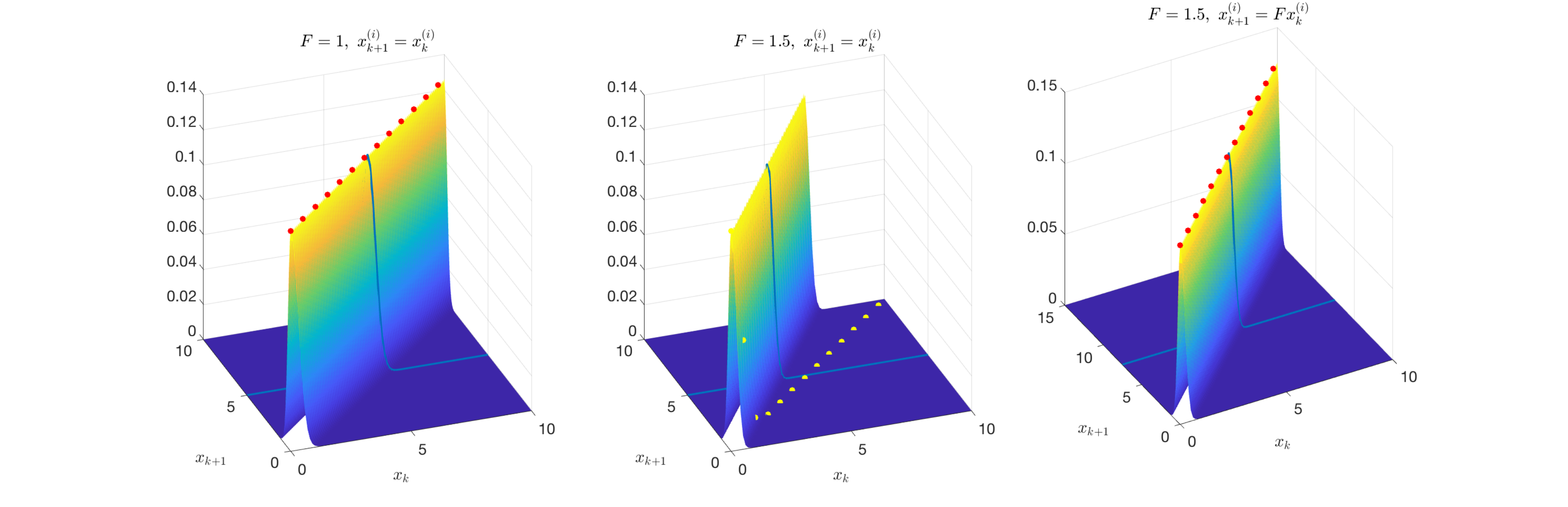}
		\caption{Transition probability matrices - First graph for random walk model and no grid movement, second for a model with linear dynamics and no grid movement, and third for linear dynamics and proposed grid movement.}
		\label{fig:TPM}
	\end{figure*}

	To illustrate the proposed concept intuitively, let a scalar DD model \eqref{eq:asx} be considered with
	\begin{align}
	F &= 1,\ w  \sim \mathcal{N}(0,Q),
	\end{align}
	meaning w.r.t. \eqref{eq:newGridDD} that the ``new'' grid remains unchanged, i.e., 	
	\begin{align}
	x_{k+1}^{(i)} &= x_{k}^{(i)} \ \forall i.\label{eq:pointsEq}
	\end{align}
	Then, the $j$-th row, and $i$-th column element of the TPM matrix is constructed from the Gaussian transition kernel with the mean $x_{k}^{(i)}, \forall i,$ as
	\begin{align}
	\bfT_{j,i}  &=  \frac{1}{\sqrt{(2\pi) Q  }} \exp{\left(-\frac{1}{2 Q}(x_{k+1}^{(j)}-x_{k}^{(i)})^2 \right)}.
	\end{align}
	In this case, the reason for a diagonal structure of the TPM can be found in the fact, that the difference $x_{k+1}^{(j)}-x_{k}^{(i)}$ is always an integer multiplier of $\Delta_k$. This is illustrated in Fig. \ref{fig:TPM}, left plot.  Then, it is possible to compute and store just a single row of the TPM and convolute it with the initial PDF. 
	However, for $F\neq1$ and arbitrary grid movement, the TPM does not have a diagonal structure, and the difference $x_{k+1}^{(j)}-Fx_{k}^{(i)}$ is not a multiplier of $\Delta_k$ as can be seen in Fig. \ref{fig:TPM}, middle plot. To obtain the TPM in the diagonal structure and thus to be able to apply \eqref{eq:tpmconvFFT}, the ``new'' grid must fulfill \eqref{eq:newGridDD} as illustrated in Fig. \ref{fig:TPM}, right plot.
	
	
	For $n_x=2$, each row of TPM is, when suitably reshaped to the physical space, a representation of a two-dimensional Gaussian PDF (with a flowing mean, and the same covariance for each row).
	
	An example of the reshaping of a TPM row representing the Gaussian PDF to the physical space is
	\begin{align}
	\setlength\arraycolsep{1.5pt}
	\psi \left(
	\begin{bmatrix}
	1 & 1 & 1 & 1 & 3 & 1 & 1 & 1 & 1
	\end{bmatrix} \right)
	= \begin{bmatrix}
	1 & 1 & 1 \\
	1 & 3 & 1 \\
	1 & 1 & 1 \\
	\end{bmatrix}.
	\end{align}
	The procedure is analogous for $n_x\geq3$.

	
	
	 \subsection{Efficient PMP for CD Model} \label{sec:contDyn}
	Let the forced grid movement $\dot{\bfx}$ be given by\eqref{eq:forcMovGrid}. Then, the FPE advection part \eqref{eq:fokker} taking into account the grid movement reads \citep{LiPe:97}
	\begin{align}
	\nabla\cdot  \left(\bfA \bfx\ p_{\mathbfcal{X}(t)}(\bfx)  \right) - \nabla p_{\mathbfcal{X}(t)}(\bfx)\ \dot{\bfx}=
	\operatorname{trace}(\bfA) p_{\mathbfcal{X}(t)}(\bfx),
	\label{eq:movFPEadv}
	\end{align} 
	where $\operatorname{trace}(\bfA)$ is the trace of the matrix $\bfA$.
	
	The problematic part of the advection $ \nabla p_{\mathbfcal{X}(t)}(\bfx) \left(\bfA \bfx\right)$ \eqref{eq:fokker}, which is causing the $\bfF_{\text{diff}}(t)$ in \eqref{eq:FKEnumSol} to be time-dependent was disposed of by using - in a sense - a Lagrangian approach \cite[pp.~7]{Ma:19}.
	The FPE to be solved then reads
	\begin{align}
	{\pderiv{p_{\mathbfcal{X}(t)}(\bfx)}{\bfv}} &= - \operatorname{trace}(\bfA)  p_{\mathbfcal{X}(t)}(\bfx) \nonumber\\
	& +\frac{1}{2}\nabla\cdot\left(\bfQ\left(\nabla^T p_{\mathbfcal{X}(t)}(\bfx)\right) \right),\label{eq:easyFPE}
	\end{align}
	where $\bfv = \begin{bmatrix}
	-\bf{\dot{x}} & 1
	\end{bmatrix} $, and \begin{align}
		{\pderiv{p_{\mathbfcal{X}(t)}(\bfx)}{\bfv}} = \bfv \begin{bmatrix}
	\nabla p_{\mathbfcal{X}(t)}(\bfx) \ ,& \ \pderiv{p_{\mathbfcal{X}(t)}(\bfx)}{t}
	\end{bmatrix}^T.
	\end{align}

	
	Compared to the discrete model and predictor, direct calculation of the $m$-th row $\bfT_{m,:}$ of the TPM, i.e., the row calculation without the need of constructing the entire matrix $\bfT$, is not straightforward. However, it is highly desirable to compute just a single row to save memory and operations.
	 
	 First define an $n_x$-dimensional finite difference diffusion matrix \citep{LiWaYaZh:19}
	 \begin{align}
	 	\bfF_{\text{diff}}(t) = \bfI - \Delta t \operatorname{trace}(\bfA) + \bfS_1 \otimes ... \otimes \bfI + ... + \bfI \otimes ... \otimes  \bfS_{n_x},
	 \end{align}
	where $\bfI \in \real^{N \times N}$, and the tridiagonal matrix is
	\begin{align}
		\bfS_i = \frac{\Delta t}{\bfDelta^2_{k,i}(t)} \text{tridiag}(\frac{\bfQ_{i,i}}{2},-\bfQ_{i,i},\frac{\bfQ_{i,i}}{2}).
	\end{align}
	The $l$-th power \eqref{eq:FKEnumSol} of $\bfF_{\text{diff}}(t)$ can be calculated using eigenvalue and eigenvector form as
	\begin{align}
		\bfT_k = (\bfR \bfLambda_{t_k}\bfR^{-1})^l,\label{eq:decomp}
	\end{align}
where $\bfR$ is a matrix of eigenvectors as columns, and $\Lambda_{t_k}$ is a diagonal matrix with eigenvalues $\lambda_{t_k}$ on diagonal.. As shown further \eqref{eq:eigVal},\eqref{eq:eigVec}, thanks to the grid movement, the eigenvector matrices can be treated as constant, while the eigenvalues $\lambda$ are time-dependent (similarly to $\bfF_{\text{diff}}$), therefore the decomposition \eqref{eq:decomp} can be treated as	
	\begin{align}
	\bfT_k = \bfR \underbrace{\left(\bfLambda_{t_k+(l-1)\Delta t}\odot \cdots \odot\bfLambda_{t_k+\Delta t} \odot \bfLambda_{t_k} \right)}_{\bfLambda_{\text{pow}}}\bfR^{-1}.
	\end{align}
	 Because the matrix $\bfF_{\text{diff}}$ is Toeplitz, its eigenvalues, reshaped to physical space, are \citep{Sa:06,St:07}
	\begin{align}
	\lambda_{i,j,...} = a + 2\ b \cos \left( \frac{i\pi}{N_1+1} \right) +  2\ c \cos \left( \frac{i\pi}{N_2+1}\right) +..., \label{eq:eigVal}
	\end{align}
	$a = 1- \Delta t \operatorname{trace}(\bfA) - \frac{\bfQ_{1,1}\Delta t}{\Delta_{k,1}^2(t)} - \frac{\bfQ_{2,2}\Delta t}{\Delta_{k,2}^2(t)}- ...$, $b =  \frac{\bfQ_{1,1}\ \Delta t}{2\Delta_{k,1}^2(t) }$, $c =  \frac{\bfQ_{2,2}\ \Delta t}{2\Delta_{k,2}^2(t) }$, and etc.
	The corresponding $i,j,...$ eigenvector's (in physical space) $u,k,...$-th component\footnote{	To put the eigenvectors into a standard form the inverse physical space transformation $\psi$ can be used
			$\psi^{-1}(\bfr_{i,j,...})$.} 
\begin{align}
	r_{i,j,...}^{(k,u,...)} = \sin\left( \frac{ik\pi}{N_1+1} \right)  \sin\left( \frac{ju\pi}{N_2+1} \right) ...,
	\\i,k = 1,...,N_1 \ ,\ j,u= 1,...,N_2. \label{eq:eigVec}
	\end{align}
	For the eigenvector matrix, it also holds that 
	\begin{align}
	\bfR^{-1} = \frac{2}{N+1} \bfR.
	\end{align}

			It can be seen that a formula to calculate an arbitrary element of $T_{j,i}$ could be derived, and the FFT-based convolution could be used analogously to the discrete case. However, thanks to the special form of the eigenvalue matrix, fast sine transform $\mathcal{S}$ can be used instead \citep{St:07} as
	 $	\widetilde P_{{k+1}}^{(:)} = \mathcal{S} \left( \widetilde{\diag(\bfLambda_{\text{pow}})} \odot \mathcal{S}(\widetilde{ P}_{{k}}^{(:)}) \right)$, which leads to a more efficient implementation.

	\subsection{Accuracy of efficient PMP}
	The proposed efficient PMP algorithms do not use any additional (w.r.t. Standard PMP) approximation and the results are as accurate as with standard PMPs. It could be argued that an error can arise when using interpolation, but when linearly interpolating PMD, no information should be lost.
	
	\section{Numerical Illustration}
	The proposed efficient PMP (ePMP) is compared with the standard PMP in two scenarios differing in state dimension $n_x$ and number of points $N_{pa}$ used, namely
	\begin{itemize}
		\item Scenario 1: $n_x=2$, $N_{pa}=99$,
		\item Scenario 2: $n_x=5$, $N_{pa}=8$.
	\end{itemize}
	The ePMP and PMP for a DD model provide the same estimates (i.e., the same estimation performance) with different computational complexity. Thus, in Table \ref{tab:D}, the computational complexity is summarized only, where the massive complexity reduction of the ePMF can be seen. 
	The numerical illustrations were computed in MATLAB\textregistered\ 2021b. 
	 The FFT implementation, available in \citep{Lu:22}, was used to calculate convolution \eqref{eq:tpmconvFFT}.
	

	\begin{table}[]
		\centering
		\caption{One step prediction time in seconds.}\label{tab:D}
		\begin{tabular}{c|cc}
			$N_{pa} = 99, n_x = 2$           & Computational time   \\\hline
			Standard PMP   &   1.4644                    \\
			Efficient PMP   &    0.0034374             \\ 
			$N_{pa} = 8, n_x = 5$ & Computational time   \\\hline
			Standard PMP  & 22.412          \\
			Efficient PMP  & 0.029077  
		\end{tabular}
	\end{table}

	
	\section{Concluding Remarks}
	The paper dealt with state prediction by the point-mass method numerically solving the CKE or the FPE. In particular, the stress was laid on computationally efficient predictive PDF calculation. The proposed computationally and memory-efficient approach is based on the smart design of a predictive grid of points allowing computation of only a row of the transition probability matrix. The computed matrix row then enables the usage of the FFT-based prediction PDF calculation. The proposed efficient PMP reduces the \textit{quadratic} computational complexity of the standard PMP into the \textit{log-linear} complexity with regard to the number of grid points. The theoretical results were verified in a numerical study. Future research will focus on a grid design and ePMP application in a complete filter design for a navigation task.
	
	\bibliography{literatura}
\end{document}